\documentclass{aa}

\usepackage{graphicx}
\usepackage{txfonts}
\defcitealias{hqmond}{HQ87}

\begin{document} 

\title{Testing MOND gravity in the shell galaxy NGC\,3923}
%\subtitle{ }

\author{M. B\'{i}lek\inst{1}\fnmsep\inst{2}
\and
B. Jungwiert\inst{1}\fnmsep\inst{2}
\and 
L. J\'{i}lkov\'{a}\inst{3}
\and
I. Ebrov\'{a}\inst{1}\fnmsep\inst{2}
\and
K. Barto\v{s}kov\'{a}\inst{1}\fnmsep\inst{3}
\and
M. K\v{r}\'{i}\v{z}ek\inst{1}\fnmsep\inst{4}
}

\institute{Astronomical Institute, Academy of Sciences of the Czech Republic, Bo\v{c}n\'{i} II 1401/1a, CZ-141 00 Prague, Czech Republic\\
\email{michal.bilek@asu.cas.cz}
\and
Charles University in Prague, Faculty of Mathematics and Physics, Ke~Karlovu 3, CZ-121 16 Prague, Czech Republic
\and 
Department of Theoretical Physics and Astrophysics, Faculty of Science, Masaryk University, Kotl\'a\v{r}sk\'a 2, CZ-611\,37 Brno, Czech Republic
\and
Astronomical Institute, Faculty of Mathematics and Physics, Charles University in Prague, V Hole\v{s}ovi\v{c}k\'{a}ch 2, CZ-180 00 Prague, Czech Republic
}

\date{Received May 00, 2013; accepted June 00, 2013}

%-------------------------------------------------------------------

\abstract
{The elliptical galaxy NGC\,3923 is surrounded by numerous stellar shells that are concentric arcs centered on the galactic core. They are very likely a result of a minor merger and they consist of stars in nearly radial orbits. For a given potential, the shell radii at a given time after the merger can be calculated and compared to observations. The Modified Newtonian Dynamics (MOND)  is a theory that aims to solve the missing mass problem by modifying the laws of classical dynamics in the limit of small accelerations. \citet{hq87} claimed that the shell distribution of NGC\,3923  contradicted MOND, but \citet{milgsh} found several substantial insufficiencies in their work. }
{We test whether the observed shell distribution in NGC\,3923 is consistent with MOND using the current observational knowledge of the shell number and positions and of the host galaxy surface brightness profile, which supersede the data available in the 1980s when the last (and negative) tests of MOND viability were performed on NGC\,3923.
  }
{Using the 3.6\,$\mu$m bandpass image of NGC\,3923 from the Spitzer space telescope we construct the mass profile of the galaxy. The evolution of shell radii in MOND is then computed using analytical formulae. We use 27 currently observed shells and allow for their multi-generation formation, unlike
the Hernquist \& Quinn one-generation model that used the 18 shells known at the time.
}
{Our model reproduces the observed shell radii with a maximum deviation of $\sim$\,5\% for 25 out of 27 known shells while keeping a reasonable formation scenario. A multi-generation nature of the shell system, resulting from successive passages of the surviving core of the tidally disrupted dwarf galaxy, is one of key ingredients of our scenario supported by the extreme shell radial range. The 25 reproduced shells are interpreted as belonging to three generations.
 }
%{???? MOND gravity passes another test.}
{}

\keywords{
Gravitation --
Methods: analytical --
Galaxies: elliptical and lenticular, cD --
Galaxies: formation --
Galaxies: individual: NGC\,3923 --
Galaxies: kinematics and dynamics
}

\maketitle

%-------------------------------------------------------------------

\section{Introduction} \label{sec:intro}

\subsection{Modified Newtonian Dynamics}
Numerous observations suggest that the magnitude of acceleration of a body, $a$, moving in a gravitational field is often tightly related to the magnitude of the Newtonian gravitational acceleration produced by the known forms of matter, $a_\mathrm{N}$ \citep{famaey12}. This relation says that $a=a_\mathrm{N}$ if $a_\mathrm{N}\gg a_0$, and $a = \sqrt{a_\mathrm{N} a_0}$ if $a_\mathrm{N}\ll a_0$, where $a_0\approx 10^{-10}$\,m\,s$^{-2}$ \citep{milg83b, milg83c, milg83a}. The existence of this relation is unexpected in the universe governed by general relativity and dark matter. In contrast with the dark matter, the baryons also are transported by non-gravitational processes (supernova explosions, gas cloud collisions, matter transport by AGN, etc.).  Moreover, the amount of baryonic matter is often negligible in comparison with dark matter. Therefore, it is surprising when the baryons alone seem to determine the dynamics of the system. This can be clearly seen on the example of the outskirts of disk galaxies, where the rotational curves of molecular gas can be  predicted, including small details, from the distribution of the extremely rarefied baryonic matter. This behavior is the motivation for the  hypothesis of Modified Newtonian Dynamics (MOND) \citep{milg83a}. It tries to explain the missing mass problem by a modification of the laws of the Newtonian dynamics for accelerations lower than $a_0$. In this way, it eliminates or greatly reduces the need to postulate the existence of the dark matter.

Several MOND theories have been built. Generally, a~MOND theory has to reduce to conventional Newtonian dynamics  in the  limit of $a_0\rightarrow 0$ (similarly as the quantum theory reduces to classical physics for $\hbar\rightarrow 0$). In the opposite limit $a_0\rightarrow \infty$, $G\rightarrow 0$ with $Ga_0$ fixed, the space-time scaling symmetry emerges. For a system consisting of massive particles $m_i$, $i=1,2,\dots ,N$, it means that if the equations of the theory are valid for the trajectories $\vec{r}_i(t)$, then they also hold for the trajectories $\lambda\vec{r}_i(\lambda t)$  \citep{milg09}. It can be shown that this behavior is an equivalent description of the law $a=\sqrt{a_0 a_N}$ for a point mass moving in a spherically symmetric gravitational field. A list of many MOND theories can be found in the thorough review of MOND by \citet{famaey12}. The theoretical motivation for this modification is still missing. It is unclear whether MOND is a modification of gravity or a modification of inertia. In the first case, the acceleration of a body depends only on the instantaneous value of the gravitational acceleration at its position; in the second case it generally depends on the full trajectory of the body \citep{milgrom06}. 

To avoid the necessity of solving the partial differential equations of the fully fledged MOND theories, the algebraic relation  $a\mu\left(\frac{a}{a_0}\right)=a_\mathrm{N}$ \citep{milg83a} is employed frequently. Here $\mu(x)$ is the interpolating function satisfying $\mu(x)\approx 1$ for $x\gg 1$, and $\mu(x)\approx x$ for $x\ll 1$. In the known modified gravity theories, the algebraic relation is precise for test particles moving in the gravitational field of a spherically symmetric matter distribution and several other highly symmetric types of systems (see, e.g., \citealp{brada94}).  However, it is well known it cannot constitute a general law of dynamics; for example one can easily prove the momentum is not conserved for an isolated system made of two particles with different masses.

In spite of the missing theoretical grounds, MOND was remarkably successful on the scale of groups of galaxies and smaller. It naturally explains  the baryonic Tully-Fisher \citep{milg83b, mcgaugh11} and Faber-Jackson \citep{milg83b, sanders2010} relations. The modified dynamics gained respect thanks to its ability to predict rotational curves of disk galaxies on the basis of their visible mass distribution (see, e.g., Figs.~21-27 in \citealp{famaey12}), even under strong observational constraints  \citep{angus13}. Ellipticals are often embedded  in gravitational fields that are too strong for the MOND effects to show up. The rare exceptions are NGC\,720 and NGC\,1521 in which \citet{milgrom12xray} verified the MOND expectations on the distribution of X-ray gas.  The modified dynamics successfully explains the rotational curves of tidal dwarf galaxies, which is a problematic task for the dark matter theory \citep{gentile07}. The  MONDian numerical self-consistent simulations of individual disk galaxies as well as interacting galaxies by \citet{tiret08} show the observed morphology. 
Another example is the high collision velocity of the Bullet Cluster, which is much higher than the escape speed predicted by the classical dynamics with dark matter. On the contrary, this velocity is not extraordinary for colliding galactic clusters in the MOND cosmological simulations \citep{angus11}. The escape velocity from the Milky Way is compatible with the MOND prediction \citep{wu08}. According to \citet{lughausen13}, MOND reproduces the qualitative properties of the rotational curves of polar ring galaxies. MOND predictions have also been confirmed for gravitational lensing by isolated galaxies \citep{milgrom13}. 

On the other hand, MOND itself does not suffice to explain the missing mass problem on the scales of galaxy clusters and higher. A possible solution is to add hot dark matter in the form of sterile neutrinos \citep{angus08, famaey08}. These problems could disappear after we find the final MOND theory.

\subsection{Shell galaxies}
\begin{figure}
\sidecaption
\resizebox{\hsize}{!}{\includegraphics{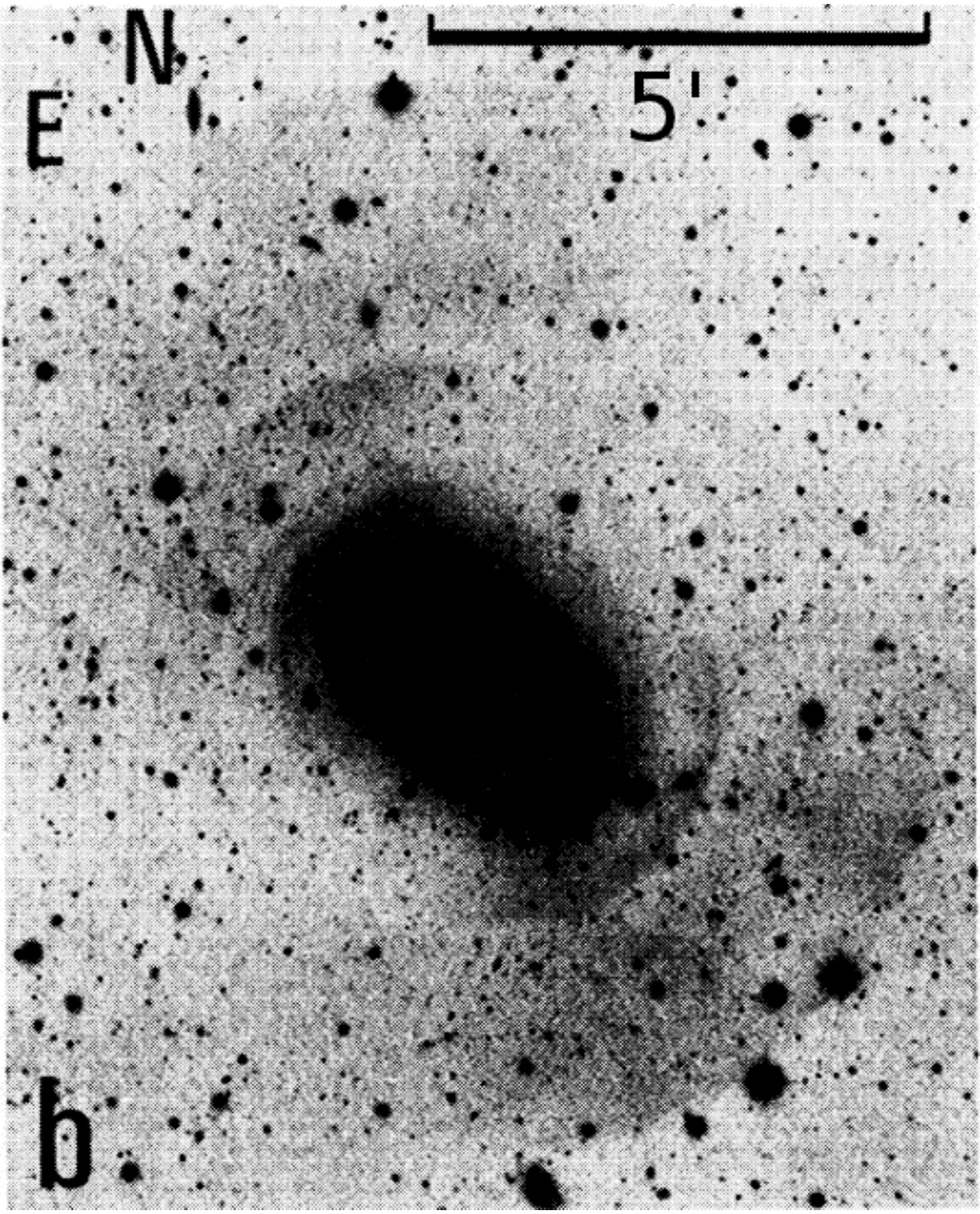}} % one column fig.
\caption{NGC\,3923 is an example of a Type I shell galaxy and the object of our study. The image comes from \citet{malin83}.}
\label{fig:fig}
\end{figure}

Shells in galaxies are arc-like sharp-edged faintly-glowing features, which are centered on the galactic core (Fig. \ref{fig:fig}). The shells are widely believed to be remnants of minor galactic mergers. The formation scenario was first proposed by \citet{quinn83} and further discussed by \citet{quinn84} and others. A~small low-mass galaxy, the secondary, encounters a much bigger and more massive galaxy, the primary. When the secondary gets close to the center of the primary, it is partly or totally disrupted by tidal forces. The stripped stars start to freely move in the potential well of the primary. When they reach the apocenters of their orbits, they slow down and form kinematical density waves which are observed as the shells. If the core of the secondary survives the first passage through the primary galaxy, it is decelerated by dynamical friction. It can start oscillating in the  potential well of the primary. Each time, it goes through the primary center it loses part of its stars and so gives rise to another generation of shells. Typically, the luminosity of the shells constitutes small percentage of the host galaxy luminosity (e.g., \citealp{prieur88, DC86}). Thus we judge that the secondary had to have mass of several percent of the mass of the primary.

A special kind of shell galaxy, the Type\,I \citep{wilkinson87}, is characterized by its regularity (see Fig. \ref{fig:fig}). The shells form an axially symmetric structure, the axis of which coincides with the major axis of the host galaxy. The shells are confined by a double cone centered at the galactic core. When sorted according to the distance from the galaxy center, almost every shell lies on the opposite side of the galaxy from its immediate predecessor and successor. This property is called the interleaving in radius. Type I shell galaxies most likely result from nearly radial mergers along the  major axis of the primary \citep{hq88, hq89}.

Type\,I shell systems can be useful for constraining the potential of the host galaxy on the basis of the positions of the shells \citep{hq87}, because a relation exists that allows the shell radii to be calculated for a given potential at a given moment after the merger.  To derive this relation, we can restrict ourselves to particles moving along the major axis of the galaxy, since the edges of shells form surfaces close to spherical caps. To proceed analytically, a simplifying assumption has to be introduced.  All the stars that form the shells are released from the primary center  at the same moment and their velocity vectors point in the same direction. In other words,  the considered stars differ only by their velocity modulus at the moment of they are released. 
The shell edges are made of stars reaching their apocenters. The  number of a shell, $n=0,1,2,\ldots$, is the number of complete oscillations performed by these stars  since the moment they were released. An oscillation is defined as the movement between two subsequent apocenters;  for example, the zeroth shell is made of stars near their first apocenter because they are just finishing one half of an oscillation. To a first approximation, at a time $t$ after the decay of the secondary, the radius of a shell can be calculated as the radius at which stars just reach their apocenter \citep{quinn84}.  The $n$-th shell is thus approximately located at distance $r_{\mathrm{A},n}$ from the center of the primary satisfying the equation
\begin{equation}
t = P(r_{\mathrm{A},n})(n+1/2),  
\label{eq:pos}
\end{equation}
where $P(r)$ denotes the period of radial oscillation at the galactocentric radius $r$ (i.e.,  2\,$\times$\,the free fall time). We note that the shells with high numbers are located at smaller galactocentric radii than those with lower numbers and that shells increase their radii with time. The value of $P(r)$ can be calculated from the primary potential $\phi$ by the relation
\begin{equation}
P(r) = \sqrt{2}\int_0^r\left[\phi(r)-\phi(x)\right]^{-1/2}\mathrm{d}x.
\label{eq:per}
\end{equation}
The positions of shells calculated by the relations~(\ref{eq:pos}) and (\ref{eq:per}) are in good agreement with our test-particle simulations (see \citealp{ebrova12}, Sect.~6, for their description). A further improvement can be reached by performing a correction. The conservation of energy requires for each particle that
\begin{equation}
r_{\mathrm{S},n} = \phi^{-1}\left[\phi(r_{\mathrm{A},n})-\frac{1}{2}v_{\mathrm{S},n}^2\right],
\label{eq:cor}
\end{equation}
where $r_{\mathrm{S},n}$ is the corrected radius of the $n$-th shell, $v_{\mathrm{S},n}$ the velocity of the particles  on its edge,  and $\phi^{-1}$ the inverse of $\phi$. We do not offer the analytical expression for $v_{\mathrm{S},n}$. However, it can be approximated by the phase velocity of apocenters $v_{\mathrm{A}, n}$ because the two velocities are very similar (see \citealp{ebrova12}, Table~1). The derivative of Eq.~\eqref{eq:pos} with respect to $t$ yields
\begin{equation}
v_{\mathrm{S},n}\approx v_{\mathrm{A},n} = \frac{1}{\left(n+1/2\right)\frac{\mathrm{d}P(r)}{\rm{d}r}\big{\vert}_{r=r_{\mathrm{A},n}}}.
\label{eq:vs}
\end{equation}
Using test-particle simulations, we have verified that this approximation works almost precisely. The simulations show that the shells with odd numbers tend to appear on the side of the primary galaxy that lies in the direction of the arrival of the secondary while the even shells appear on the opposite side. To take this property into account, we will hereafter take $r_{\mathrm{S},n}$ as positive for even $n$ and as negative for odd $n$.

Self-consistent simulations show that formation of shells in several generations is common \citep{cooper11,bartoskovaselfcon,SegDup}. For the minor merger model, this manner of formation is actually needed to explain the high radial ranges observed in shell galaxies (the radial range is the ratio of radii of the innermost and the outermost shell); for example, in our test-particle simulations \citep{ebrova12} and that of \citet{DC86}, where the shells are forced to form in a single generation, more than ten shells are never visible. If the secondary core survives a passage through the primary center, it loses part of its kinetic energy by dynamical friction, so the shells formed in the subsequent generation lie in lower range of radii. 

Shell formation is a complex process. The disruption of the secondary is actually a gradual rather than an instantaneous event. The secondary is  also affected by dynamical friction during its decay. These processes can have an impact on the distribution of shells \citep{DC86,ebrovagraddecay}. However, the self-consistent simulations of \citet{bartoskovaselfcon}, Fig.~1, show that these effects cause a shift of the shells of less than 10\%. In MOND, furthermore,  the dynamical friction tends to be lower than in the analogous system with dark matter \citep{tiret08}. On the other hand, the disruption of the secondary in MOND is probably more gradual than in the classical dynamics because of the external field effect \citep{milgsh}.

\subsection{NGC\,3923 and MOND}
The galaxy NGC\,3923, the main object of interest in this study, is an E4--5 Type\,I shell galaxy in the Hydra constellation. It is a record object among shell galaxies for the number of shells surrounding it: 27 shells have been found so far. It hosts the greatest known shell in the Universe with a radius of $\sim$\,130\,kpc. It is also exceptional for the radial range of its shells which is 65. All of which makes NGC\,3923 the most studied shell galaxy. 

\citet{hqmond} reported, that the distribution of shells in NGC\,3923 contradicts MOND. Modeling the  galaxy with a de~Vaucouleurs sphere, they derived the relation between the shell numbers and their radii using Eqs. \eqref{eq:pos} and \eqref{eq:per}. Their method required that the shell system be formed in a single generation and that the all existing shells be observed except of those father from the galactic center than the biggest known shell. However, \citet{milgsh} found several substantial deficiencies in their analysis, some of which concerned the assumptions of Hernquist and Quinn, which were too strong and probably incorrect. Others pointed to their method, because it can lead to incorrect conclusions if Eqs. \eqref{eq:pos} and \eqref{eq:per} do not work precisely. The method is also sensitive to observational errors in the shell radii. To our knowledge, shell galaxies have not been investigated in the context of MOND since that time. At the time when \citet{hqmond} published their paper, only 18 shells were known. Further observations discovered new shells (\citealp{prieur88} and \citealp{sikkema07}), so that we know 27 shells around NGC\,3923 today.

In this paper we test whether the shell system of NGC\,3923 is consistent with the prediction of MOND gravity. To this aim, we calculated the radii of shells for this galaxy expected in MOND. The paper is organized as follows. In Sect.~\ref{sec:indata} we describe the observational data used for the calculations (shell radii and characteristics of NGC\,3923). In Sect.~\ref{sec:pot_model} we give a description of the gravitational potential model used for NGC\,3923. Section~\ref{sec:mond} includes description and analysis of the shell radii calculations.  And finally, we discuss and summarize our results in Sects.~\ref{sec:disc} and~\ref{sec:sum}, respectively.
%-------------------------------------------------------------------

\section{Input data}
\label{sec:indata}
We use the set of shell positions given in the second column of Table~\ref{tab:shpos}. The shells with the ``$+$'' sign lie on the northern side of the galaxy and those with ``$-$'' on the southern. The data for shell distances greater than $130\arcsec$ are taken from \citet{prieur88}, the rest from \citet{sikkema07}. It is not easy to say where the edge of a shell for an observed galaxy lies exactly, because it is always blurry (see, e.g., Figs.~3 and D.4 in \citealp{sikkema07}). The observers measured the position of the maximum of brightness, which does not necessarily need to be the same as the position of the edge. This uncertainty of radii of the shells reaches several percent. The shell edges form almost spherical caps. Therefore, the three-dimensional radius of a shell must be the same as the projected one.  

\begin{table}
\caption{Shells of NGC\,3923}\label{tab:shpos}
\centering
\begin{tabular}{crrrrrr}
\hline\hline
Label & $d$ $[\arcsec]$ & $G$ & $n$ & $o$ & $d_{\mathrm{model}}$ $[\arcsec]$ & $\Delta$ [\%] \\
\hline
a & $+1170$ & I & $2$ & $-1$ & $+1178$ & $0.7$ \\
b & $-840$ & I & $3$ & $-1$ & $-845$ & $0.6$ \\
c & $+630$ & I & $4$ & $-1$ & $+658$ & $4.5$ \\
d & $-520$ & I & $5$ & $-1$ & $-539$ & $3.6$ \\
e & $+365$ & I & $8$ & $-1$ & $+349$ & $4.5$ \\
f & $-280$ & II & $2$ & $1$ & $-275$ & $1.8$ \\
g & $+203$ & II & $3$ & $1$ & $+198$ & $2.5$ \\

h & $-148.5$ & II & $4$ & $1$ & $-155.2$ & $4.5$ \\
i & $+147.3$ & II & $4$ & $1$ & $+155.2$ & $5.1$ \\
j & $+128.1$ & II & $5$ & $1$ & $+127.6$ & $0.4$ \\
k & $-103.6$ & II & $6$ & $1$ & $-108.4$ & $4.6$ \\
l & $+99.9$ & II & $7$ & $1$ & $+94.5$ & $5.4$ \\
m & $-79.6$ & II & $8$ & $1$ & $-83.9$ & $5.4$ \\
n & $+72.8$ & II & $9$ & $1$ & $+75.4$ & $3.6$ \\
o & $-67.0$ & II & $10$ & $1$ & $-68.6$ & $2.3$ \\
p & $+64.1$ & III & $6$ & $-1$ & $+64.0$ & $0.1$ \\
q & $+60.4$ & II & $11$ & $1$ & $+62.9$ & $4.1$ \\
r & $-55.5$ & III & $7$ & $-1$ & $-55.9$ & $0.7$ \\
s & $+51.2$ & III & $8$ & $-1$ & $+49.7$ & $3.0$ \\
t & $-44.0$ & III & $9$ & $-1$ & $-44.8$ & $1.8$ \\
u & $+41.5$ & III & $10$ & $-1$ & $+40.8$ & $1.6$ \\
v & $-37.7$ & III & $11$ & $-1$ & $-37.5$ & $0.5$ \\
w & $+34.3$ & III & $12$ & $-1$ & $+34.7$ & $1.1$ \\
x & $+29.3$ & III & $14$ & $-1$ & $+30.2$ & $3.0$ \\
y & $-28.7$ & III & $15$ & $-1$ & $-28.3$ & $1.3$ \\
A & $+19.4$ & IV & ? & $1$ & ? & ? \\
B & $-18.0$ & IV & ? & $1$ & ? & ? \\\hline
\end{tabular}
\label{tab:sh}
\tablefoot{$d$\,--\,the observed distance of the shell from the center of NGC\,3923, data taken from \citet{prieur88} and \citet{sikkema07}. The plus sign means that the shell is situated on the northern side of the galaxy and minus on the southern; $n$\,--\,identified shell number; $G$\,--\,identified shell generation; $o$\,--\, the sign of generation; $d_{\mathrm{model}}$\,--\,modeled distance of the shell; $\Delta$\,--\,relative difference between observed and modeled distance.}
\end{table}

There are five direct distance measurements of NGC\,3923 from Earth on NED (The NASA/IPAC Extragalactic Database)\footnote{\url{http://ned.ipac.caltech.edu/}} that are all  based on the surface brightness fluctuation method. The distance varies from 19.9 to 24.0\,Mpc with the median of 22.9\,Mpc. In the following, we will use the value of 23\,Mpc.

Our goal will be to calculate the shell radii time evolution in NGC\,3923 using Eqs.~\eqref{eq:pos}--\eqref{eq:vs} in MOND. This requires a model of the mass distribution of the galaxy, which we derive by deprojecting the observed radial surface brightness profile while assuming a constant mass-to-light ratio, $M/L$.

\begin{figure}
\sidecaption
\resizebox{\hsize}{!}{\includegraphics{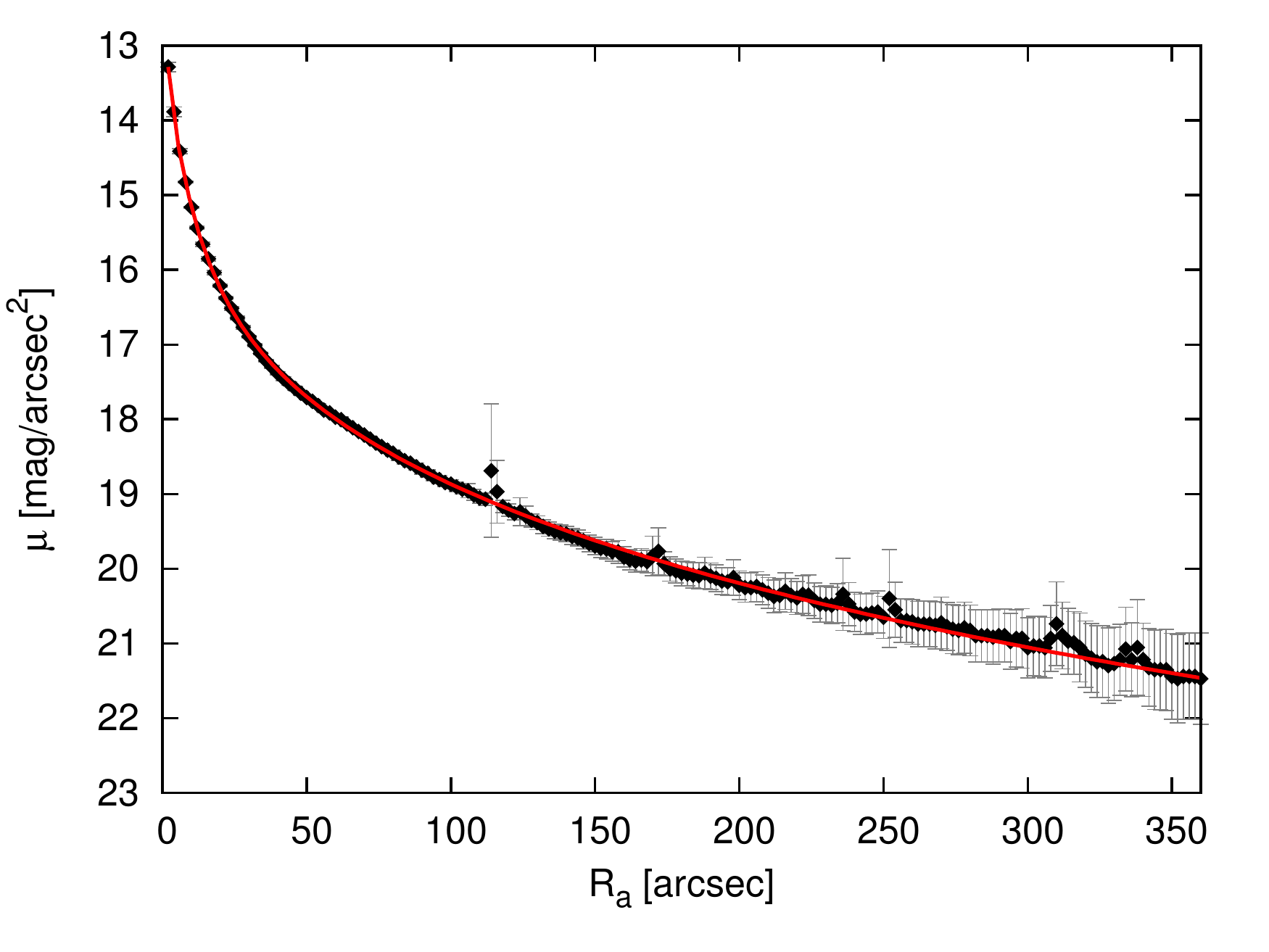}} % one column fig.
%\resizebox{\hsize}{!}{\includegraphics{fig_profile.eps}} include .eps into the final version, but .pdf to the draft versions (eps->pdf by epstopdf --autorotate=All fig_*.eps
\caption{Surface brightness profile of NGC\,3923 (Spitzer Space telescope IRAC 3.6\,$\mu$m bandpass). The semi-major axis of the isophotal ellipses, $R_{a}$, is given on the horizontal axis. Black points with gray error-bars show the observed profile; the red line shows fit by the sum of two S\'ersic profiles (see Table~\ref{tab:ser2}).}
\label{fig:profile}
\end{figure}

%\subsection{Surface photometry and radial profile}
To derive the radial surface brightness profile, we used a flux-calibrated image of NGC\,3923 from the Spitzer Space Telescope archive. The image is a mosaic made from 120 individual frames obtained with the InfraRed Array Camera (IRAC, see \citealp{fazio04}) in the 3.6\,$\mu$m bandpass. First, we manually masked bright foreground stars and also the brightest shells that could influence the surface photometry. The surface brightness analysis was done following \citet[][see also \citealp{gildepaz05}, and \citealp{munoz-mateos09}]{munoz-mateos13}. Galaxy isophotes were fitted using the IRAF \citep{iraf} %\footnote{IRAF is the Image Reduction and Analysis Facility, a general purpose software system for the reduction and analysis of astronomical data. IRAF is written and distributed by the National Optical Astronomy Observatories, which are operated by the Association of Universities for Research in Astronomy, Inc., under cooperative agreement with the National Science Foundation.}
 task \texttt{ellipse}. The galaxy center was obtained using the IRAF task \texttt{centroid} and the center was kept fixed during the isophote fitting. The first run of the isophote fitting was carried out with a linear step of $6\arcsec$ in the semi-major axis. The mean ellipticity of the galaxy $(0.316 \pm 0.007)$ defined as $1-a/b$, where $a$ and $b$ are the semi-major and semi-minor axes, respectively, and the mean position angle $(48.2 \pm 0.1)\degr$ (measured counterclockwise from the north) were then obtained as intensity weighted means of individual isophotal ellipse parameters. Then we performed a second run of the \texttt{ellipse} isophotal analysis, with a linear step of $2\arcsec$ and the position angle and ellipticity fixed to the mean values. Output of this second run was used to construct the surface brightness radial profile.

The sky intensity and uncertainties in surface brightness were obtained following the method of \citet{gildepaz05} and \citet{munoz-mateos09}. To obtain the sky level, we measured the intensity in ten square regions (with the size of 40$\times$40\,pixels) randomly situated around the galaxy, far enough from its center to avoid contamination by the galaxy itself. The sky intensity was obtained as the mean value in all ten regions. The uncertainty in surface brightness is calculated as a combination of the uncertainty of intensity measured along individual isophotes (part of the \texttt{ellipse} task output) and uncertainty of the sky intensity measurement (which has two components: Poisson noise in the sky intensity and possible large-scale variations due to flat-fielding errors). Following \citet{munoz-mateos13}, we also applied a correction of the photometric calibration of IRAC for the extended source photometry. This corrects the surface brightness measured along an isophote for the extended emission of the IRAC PSF and diffuse scattering on the detector. Finally, the surface magnitude was calculated using the IRAC zero-point of 18.8\,mag (according to the IRAC Instrument Handbook\footnote{\url{http://irsa.ipac.caltech.edu/data/SPITZER/docs/irac/iracinstrumenthandbook/25/}}).

The resulting surface brightness radial profile is shown in Fig.~\ref{fig:profile}. In order to obtain a model of mass distribution of the galaxy, we fitted the surface brightness by sum of two S\'ersic profiles. Parameters of the fit are listed in Table~\ref{tab:ser2}. The fit is also shown in Fig.~\ref{fig:profile} by the red line.

\begin{table}
\caption{A two-component S\'ersic fit of the Spitzer IRAC image of NGC\,3923 in the 3.6\,$\mu$m bandpass.}\label{tab:ser2}
\centering
\begin{tabular}{ccccc}
\hline\hline
Component & $\mu_{\mathrm{e}}$ & $r_{\mathrm{e}}$ & $n$ \\
          & [mag\,arcsec$^{-2}$] & $[\arcsec]$      &     \\
\hline \\[-1.5ex]
1 &  $20.5\pm0.2$ & $235^{+20}_{-21}$       & $5.28^{+0.52}_{-0.48}$ \\[1ex]
2 &  $15.9\pm0.1$    & $7.58^{+0.32}_{-0.28}$  & $1.53^{+0.17}_{-0.13}$ \\[0.5ex]
\hline
\end{tabular}
\label{tab:fit}
\tablefoot{$\mu_{\mathrm{e}}$\,--\,effective surface brightness of the individual S\'ersic profile components;
%$m$\,--\,apparent 3.6\,$\mu$m magnitude; 
$r_{\mathrm{e}}$\,--\,effective radius; $n$\,--\,S\'ersic index. The errors indicate the 95\% confidence interval.}
\end{table}

The $M/L$ ratio in the 3.6\,$\mu$m band was obtained using Eq.~(4) in \citet{for12}
\begin{equation}
(M/L)_{3.6} = 0.92\times 10^{1.434(J-K_{\mathrm{s}})-1.380}-0.05,
\end{equation}
where $J$ and $K_{\mathrm{s}}$ are the magnitudes in the 2MASS (Two Micron All Sky Survey) J and K$_{\mathrm{s}}$ bands \citep[see][]{2mass}. The two magnitudes were obtained by fitting the 2MASS images of the galaxy in the J and K$_{\mathrm{s}}$ bands. A sum of two S\'ersic profiles was used again. We get $J=7.44$\,mag and $K_{\mathrm{s}} = 6.51$\,mag, which implies  $(M/L)_{3.6}=0.78$ in the solar units. The absolute solar magnitude in the 3.6\,$\mu$m band is 3.24\,mag \citep{oh08}.

For these values, we get the total stellar mass of NGC\,3923 of $5.1\times 10^{11}$\,M$_\odot$, which is close to the value of $5\times 10^{11}$\,M$_\odot$ given by \citet{prieur88}. For comparison, the same author measured the effective radius of the galaxy as $105\,\arcsec$. \citet{sikkema07} determined the value of the effective radius as $39\,\arcsec$ from the narrow field-of-view image from Hubble Space Telescope.

The measurements of the mass of the hot X-ray emitting gas are available up to the radius of $500\,\arcsec$. \citet{naginomatsushita} measured the dynamical mass-to-light ratio in the K band at this radius. Their value differs only by 3\% from the one given by our model of NGC\,3923 in MOND, which we derive in Sect. \ref{sec:pot_model} from the distribution of stellar mass. According to Fig.~4 in \citet{satotawara}, NGC\,3923 contains an exceptionally low hot gas fraction among early-type galaxies. Therefore we neglect its mass. 

%-------------------------------------------------------------------

\begin{figure*}
\sidecaption
\resizebox{\hsize}{!}{\includegraphics{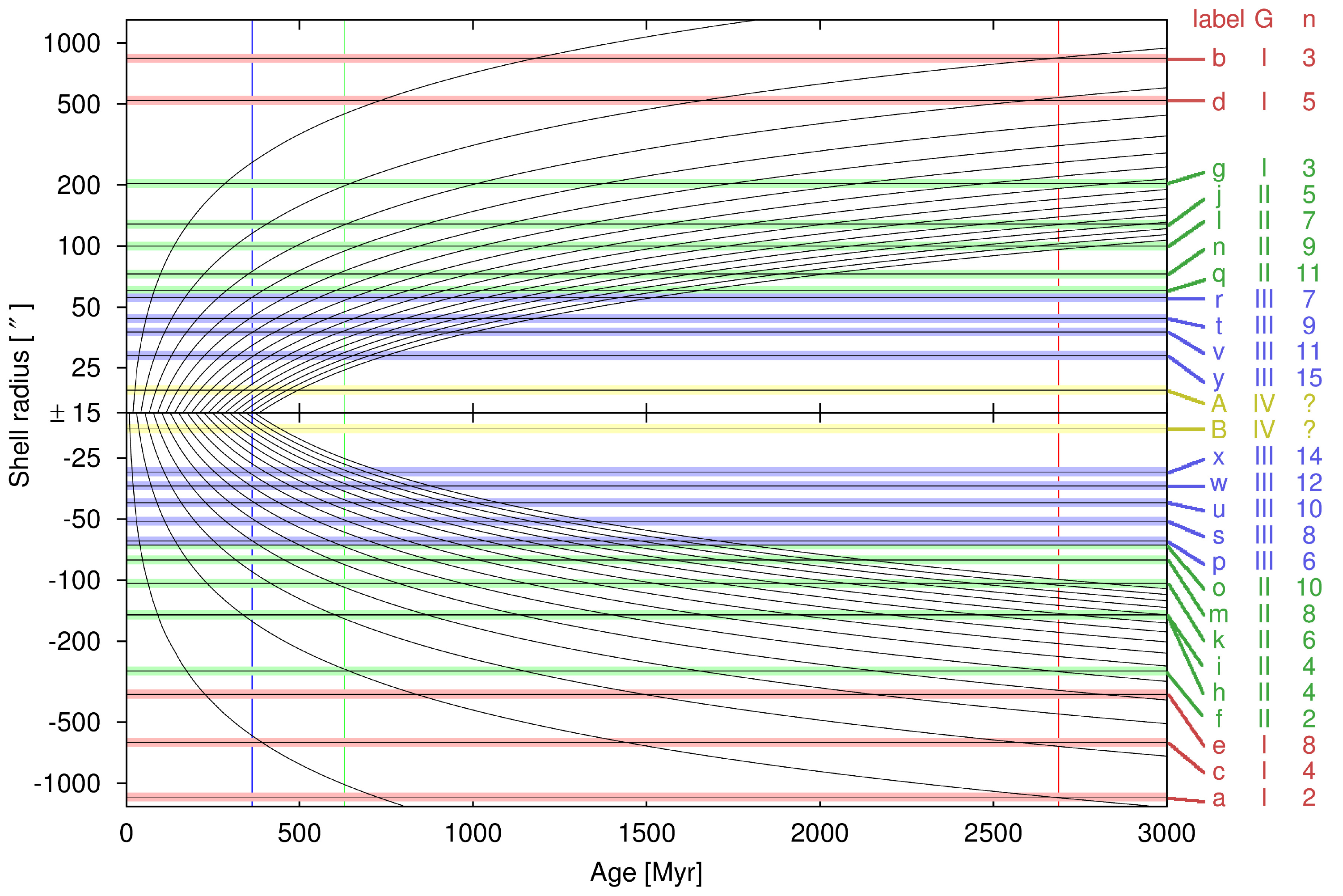}} % one column fig.
%\resizebox{\hsize}{!}{\includegraphics{fig_shcomp.eps}} include .eps into the final version, but .pdf to the draft versions (eps->pdf by epstopdf --autorotate=All fig_shcomp.eps
\caption{Calculated time evolution of shell radii in our potential (black curves).  The observed radii of shells plus 5\% uncertainty boundaries (horizontal lines). The shells from the same generations are marked by identical colors. The age of a generation is marked by the vertical line of the same color. The calculated shell radii should ideally reach their observed values at the times marked by the vertical lines. The shells can be identified in Table \ref{tab:sh} by their labels. The other two columns on the right denote the identified generation to which the shell belongs and its number. The signs of the shells from the first column of Table~\ref{tab:sh} had to be multiplied by the value in Col.  to take into account the directions of the individual collisions. The sign of shell $i$ was switched since it is interpreted as shell $h$ encircling the galaxy. The two innermost shells  probably come from the fourth generation and were not formed in the manner that our model assumes. }
\label{fig:shcomp}
\end{figure*}

\section{Model of the gravitational potential} \label{sec:pot_model}
The spatial orientation of NGC\,3923 is unknown. Our goal is only to test whether its shell distribution is consistent with MOND. Therefore, we simply assume that the galaxy is a prolate ellipsoid, the axis of which is perpendicular to the direction toward the observer. If the observer is not situated close enough to the equatorial plane of the galaxy, the shells might not be visible or they would not form  an axially symmetric structure. The prolate shape of the galaxy is also preferred because of its minor axis rotation \citep{minrot}. We repeated the test described in Sect. \ref{sec:mond} for an oblate ellipsoid and we get almost the same results.  

Deprojection of the fitted profile was performed  to obtain the volume mass density model. Subsequently, the gravitational acceleration $a_{\mathrm{N}}$ was calculated along the galactic major axis in the classical Newtonian way. The MONDian acceleration $a_{\mathrm{M}}$ was obtained from the algebraic relation \citep{milg83a}
\begin{equation}
a_{\mathrm{M}}\mu\left(\frac{a_{\mathrm{M}}}{a_0}\right) = a_{\mathrm{N}}.
\label{eq:algrel} 
\end{equation}
We choose the successful simple interpolation function \citep{famaey12}
\begin{equation}
\mu(x) = \frac{x}{x+1},
\label{eq:intfun}
\end{equation}
and the value of the Milgrom acceleration constant $a_0=1.2\times 10^{-10}$m\,s$^{-2}$.
We are aware that the use of the algebraic relation is inexact for a non-spherical galaxy. The calculations of \citet{ciotti06} suggest that this inaccuracy in acceleration is of the order of several percent.

We treat MOND as a modification of gravity, so the acceleration of a particle depends only on its immediate position. 
Integrating the MONDian acceleration along the major axis of the galaxy one gets the potential and can use  Eqs.~\eqref{eq:pos}--\eqref{eq:vs} to evaluate the shell positions at any time after releasing the stars from the secondary at the center of the primary.

%-------------------------------------------------------------------

\section{Are the shell radii of NGC\,3923 consistent with MOND?}
\label{sec:mond}
Shells in NGC\,3923, as in other shell galaxies, were probably formed in several generations. Its extreme shell radial range points to it. The progenitor of the shells was disrupted by several subsequent passages through the center of NGC\,3923, part by part at each passage. The set of shells originating from the same passage is called the generation. For the purposes of the following test, it is convenient to define the sign of the $N$-th generation: If the secondary  reached the primary center from north during its $N$-th infall, we put $o_N=1$; otherwise  $o_N=-1$. 

In order to compare the radii of the observed shells with those calculated using  Eqs. \eqref{eq:pos}--\eqref{eq:vs} one has to 1)~interpret the observed shells, i.e., to assign a shell number and generation number to each of the observed shells, and 2)~to choose the ages of the individual generations, $t_N$. Those are free parameters. If our potential is the right one, the model of shell propagation is flawless and the shell radii are measured absolutely precisely, then it is possible to identify the true interpretation, which satisfies 
\begin{equation}
r_{m,N}(t_N) = r_\lambda o_N
\label{eq:shcomp}
\end{equation}
for the all observed shells. In this equation, the observed shell labeled $\lambda$ is interpreted as the shell number $m$ from the $N$-th generation released at the time $t_N$ before the observation. The radius $r_\lambda$ is that listed in the Col. 2 of Table~\ref{tab:sh} and $r_{m,N}$ is the radius of the $m$-th shell calculated using Eqs.~\eqref{eq:pos}--\eqref{eq:vs} with the added sign. We note again that we define the sign of $r_{m,N}$ as positive for odd $m$ and negative in the other case. 

Equation \eqref{eq:shcomp} stands as the basis of our test of the potential, in which we try to find the right interpretation of the observed shells and the time passed since the individual collisions. An interpretation of the observed shells can be considered as plausible if the following criteria hold true:
\begin{itemize}
\item The relative difference $\left|\left[r_{m,N}(t_N) - r_\lambda o_N\right]/r_\lambda\right|$ is small for all the observed shells and appropriate ages of the generations.
\item  Only a few shells  escape observations. When a generation is formed, stars are released from the secondary with energies forming a continuous interval. Therefore, we expect the numbers of the observed  shells from a generation to form a series of subsequent integers. If a number is missing in this series, a shell must be escaping observations. 
If a sufficient number of shells are allowed to be missing in an interpretation, it is easy to fulfill the remaining criteria for virtually any set of shell radii.  
\item Not many generations are present. The reason is similar to that of the previous point.  
\item The $(N+1)$-th generation is younger than the $N$-th generation, for any $N$, i.e., $t_\mathrm{I}>t_\mathrm{II}>t_\mathrm{III}>\dots$
\item  The time difference between the subsequent passages of the secondary shortens because the galaxy is decelerated by dynamical friction.
\item At every infall, the secondary reaches the center of the primary from the side opposite the previous one, i.e., $o_\mathrm{I}=-o_\mathrm{II}=o_\mathrm{III}=\dots$
\end{itemize}

\noindent
If no interpretation satisfies these criteria, it means that the potential is incorrect, or that the model of shell radii evolution is too rough, or even that many not yet observed shells exist.

However, we succeeded in finding an interpretation that complies with these criteria for 25 outermost shells of NGC\,3923 very well. Maximum deviation of observed and calculated shell radii is $5.4\%$.
The identified shell numbers and generations are presented in the Cols. 3 and 4 of Table~\ref{tab:shpos}. The interpretation is also apparent from  Fig.~\ref{fig:shcomp}. In this figure, the evolution of calculated shell positions is depicted by the black curves. The observed shell radii, after the multiplication by the sign of their parent generation, are marked by the horizontal lines. The  25 outermost observed shells originate from three generations. In Fig.~\ref{fig:shcomp}, shells from the same generations are marked by the same color: red, green, and blue. The age of the generations is marked by the tree vertical lines. The calculated shell radii should ideally reach their observed values at the age of the generation. The ages of individual generations are $t_\mathrm{I}=2688$\,Myr, $t_\mathrm{II}=631$\,Myr, and $t_\mathrm{III}=364$\,Myr. The ages were chosen so that the maximum of relative deviations in each generation is minimized. According to the presented interpretation, the sign of the first generation is $o_\mathrm{I}=-1$, meaning that the secondary originally collided with the primary from south along the major axis of the primary.

We have checked that these results remain practically unchanged if the galaxy is assumed to be an oblate ellipsoid.

\section{Discussion}
\label{sec:disc}

As explained in Sect.~\ref{sec:indata}, a significant part of the deviations can be caused by observational uncertainty of the shell radii, for example, as can be seen in  Fig.~D.1 of \citet{sikkema07}, their shell no. 16 is very faint and diffuse. It was not even detected by \citet{prieur88}. This is our shell labeled $l$, which causes the largest deviation.  If we exclude this shell, the optimal age of the second generation comes out $621$\,Myr and the deviations of the second generation shells drop below $3.9\%$. 

If this interpretation of shells is correct, shells 6 and 7 of the first generation are escaping observation. This can happen because even several of the already known shells are at the limits of detectability (see Fig.~D.1 of \citealp{sikkema07}). Shells 6 and 7 should be located approximately at the radii of $-390$ and $450$ arcseconds. 

Shells can be observed only  at the distances where a sufficient number of stars have apocenters. Accordingly, every shell disappears when it grows too big. The lower the shell number, the sooner the shell vanishes. For the same reason, new shells appear at a certain minimum radius. The minimum and maximum radii are common for all the shells from the same  generation. That is why the shells with low and high numbers can be unobservable. 

There are three pairs of shells in Table~\ref{tab:shpos} that violate the interleaving in radius: $i$--$j$, $p$--$q$, and $w$--$x$. There are several possible reasons for this: 1)~some shells with radii between the pair were missed by  observations; 2)~one shell of the pair comes from a different generation from the other; or 3)~a dynamical phenomenon is present that is not included in our model. According to the presented interpretation, a missing shell must exist between shells $w$ and $x$. This is credible because there is a dust cloud visible in Fig.~D.1 of \citet{sikkema07} at the expected position of the missing shell (between shells 3 and 6 in their figure). The pair $p$--$q$ is interpreted as two shells from different generations. Shell $i$ of the remaining pair $i$--$j$ seems to be a part of shell $h$ encircling the whole galaxy, because the radii of $h$ and $i$ are very close to each other. As our unpublished simulations show (like those described in \citealp{ebrova12}), the interleaving in radius does not 
always have to be a strict law.  For some common potentials (e.g., Navarro-Frenk-White, de Vaucouleurs), shells can be formed which encircle the whole galaxy. Their radius is still well approximated by Eqs. \eqref{eq:pos}-\eqref{eq:vs} (with a deviation of a few percent). The ellipticity of the potential helps to keep the shells confined to the double cone \citep{hq89}. 

We do not present an unequivocal explanation of the origin of the two innermost shells. They probably come from the fourth generation.  Equations~\eqref{eq:pos}--\eqref{eq:vs} come from the assumption that the stars are always released from the secondary when it passes through the primary center. However, it is possible that the secondary was peeled so much by the third passage that its remnant was disrupted by the tidal force before it reached the center for the fourth time. The assumption of radiality of the stellar orbits may be too crude for these shells at smallest radii. 

At the moment, we are not able to guarantee that the shell interpretation we have presented is the only one possible. The probability with which it is possible to interpret equally well an arbitrary set of shell radii is also unknown. These questions will be the topic of our next paper.

%-------------------------------------------------------------------

\section{Summary}
\label{sec:sum}
We showed that the distribution of shell radii observed in the galaxy NGC\,3923 is consistent with MOND, at least up to the  current level of knowledge of the shell formation process. The relative deviations of shell radii in our model from observation are maximally 5.4\% for 25 out of the 27 known shells (Table \ref{tab:sh}, last column). The order of these deviations coincides with the order of the uncertainty in radii of the observed shells originating from the smoothness of their edges.  This result removes the shade of doubt shed on MOND by the work of \citet{hq87}, who state that the shell distribution of NGC\,3923 contradicts MOND. 

No tuning of potential was needed. Fixed values of the galaxy distance and mass-to-light ratio were used. They come from sources independent of the shell positions (Sect. \ref{sec:indata}).

We have assumed that MOND is the modification of gravity and that the algebraic relation between the MONDian and Newtonian accelerations holds sufficiently precisely.

The model of the mass distribution of the shell galaxy NGC\,3923 was derived on the basis of the Spitzer IRAC 3.6\,$\mu$m bandpass data. The fit parameters can be found in Table~\ref{tab:fit}. The total mass of the galaxy came out to be $5.1\times10^{11}\,\mathrm{M}_\odot$.  The MONDian gravitational potential was calculated using the algebraic relation \eqref{eq:algrel}  and the simple interpolation function \eqref{eq:intfun} along the major axis of the galaxy. The modeled shell radii were calculated using the formulas in Eqs. \eqref{eq:pos}--\eqref{eq:vs} that come from the assumption    that the shells are formed during a radial minor galactic merger and that the shells are made of stars near apocenters that move in radial orbits.

This work is the first re-investigation of the topic. It can be improved in numerous aspects. First of all, one can calculate the MONDian acceleration field using the equations of one of the full MOND theories, instead of the approximative algebraic relation. The credibility of our results, the decay of the secondary in 3-4 generations,  the age differences between them and the number of created shells, should be verified by full self-consistent simulations, like those performed by \citet{tiret08}, for example. The self-consistent simulation would also clarify the question already raised by  \citet{milgsh}: how does the external field effect influence the formation of shells? The decay of the 
secondary is expected to be more gradual in the modified than in the classical 
dynamics. The external field effect works as if the gravitational constant 
inside the secondary decreased as it moves deeper in the 
potential well of the primary.   
  
%-------------------------------------------------------------------

\begin{acknowledgements}
We acknowledge the support from the following sources: Czech support for the
long-term development of the research institution RVO67985815
(MB, BJ, IE, KB, and MK), the project SVV267301 by the Charles University
in Prague (MB, IE, and MK) and the grant MUNI/A/0735/2012 by the Masaryk
University in Brno (LJ and KB). 
%Spitzer
This work is based on observations made with the Spitzer Space Telescope, which is operated by the Jet Propulsion Laboratory, California Institute of Technology under a contract with NASA.
%2MASS
This publication makes use of data products from the Two Micron All Sky Survey, which is a joint project of the University of Massachusetts and the Infrared Processing and Analysis Center/California Institute of Technology, funded by the National Aeronautics and Space Administration and the National Science Foundation.
\end{acknowledgements}

%-------------------------------------------------------------------

\bibliographystyle{aa}
\bibliography{citace}

%-------------------------------------------------------------------
\end{document}